\documentclass[a4paper]{jpconf}
\usepackage[utf8]{inputenc}
\usepackage[dvipsnames]{xcolor}
\usepackage{amsmath,amssymb,amsthm,verbatim}
\usepackage[pdftex]{graphicx}
\usepackage[toc,page]{appendix}
\usepackage{tabularx,ragged2e,booktabs,caption, subcaption,longtable}
\usepackage{url,hyperref}
\usepackage{listings}
\usepackage{paralist}
\usepackage{enumitem}
\usepackage{latexsym}
\usepackage{amssymb}
\usepackage{amsmath}
\usepackage{graphicx}
\usepackage{colortbl}
\usepackage{bm}
\usepackage{float}
\usepackage{xr}
\usepackage{multirow}
\usepackage{afterpage}
\usepackage[printonlyused]{acronym}
\usepackage{cite}

\usepackage{caption}

\definecolor{green}{HTML}{66FF66}
\definecolor{myGreen}{HTML}{009900}

\begin{document}
 \title{Search for Light Neutral Bosons in the TREK/E36 Experiment at J-PARC}

 \author{Dongwi H. Dongwi  (on behalf of the TREK/E36 Collaboration)}

 \address{Lawrence Livermore National Laboratory, 7000 East Ave., Livermore, CA 94550-9234}

 \ead{dongwi1@llnl.gov}

 \let\clearpage\relax
 \begin{abstract}
  The Standard Model (SM) represents our best description of the subatomic world and it has been very successful in explaining how elementary particles interact under the influence of the fundamental forces. Despite its far reaching success in describing the building blocks of matter, the SM is still incomplete; falling short to explain dark matter, baryogenesis, neutrino masses and much more. The E36 experiment conducted at J-PARC in Japan, allows for sensitivity to search for light $U(1)$ gauge bosons, in the muonic $K^+$ decay channel. Such $U(1)$ bosons could be associated with dark matter or explain established muon-related anomalies such as the muon $g_{\mu}-2$ value, and the proton radius puzzle. A scintillating fiber target was used to stop a beam of positively charged $K$ mesons. The $K^+$ products were detected with a large-acceptance toroidal spectrometer capable of tracking charged particles with high resolution, combined with a large solid angle CsI(TI) photon detector and particle ID systems. A realistic simulation was employed to search for these rare decays in the mass range of 20$-$100 MeV/$c^2$. Preliminary results of the upper limits for the $A^\prime$ branching ratio $\mathcal{B}r(A^\prime)$ extracted at 95\% CL, will be discussed.
 \end{abstract}

 \include{ntroduction}
  \section{Light Neutral Boson Search with the TREK/E36 Detector Apparatus}
\begin{figure}[H]
  \begin{center}
   \centering\includegraphics[scale=.54]{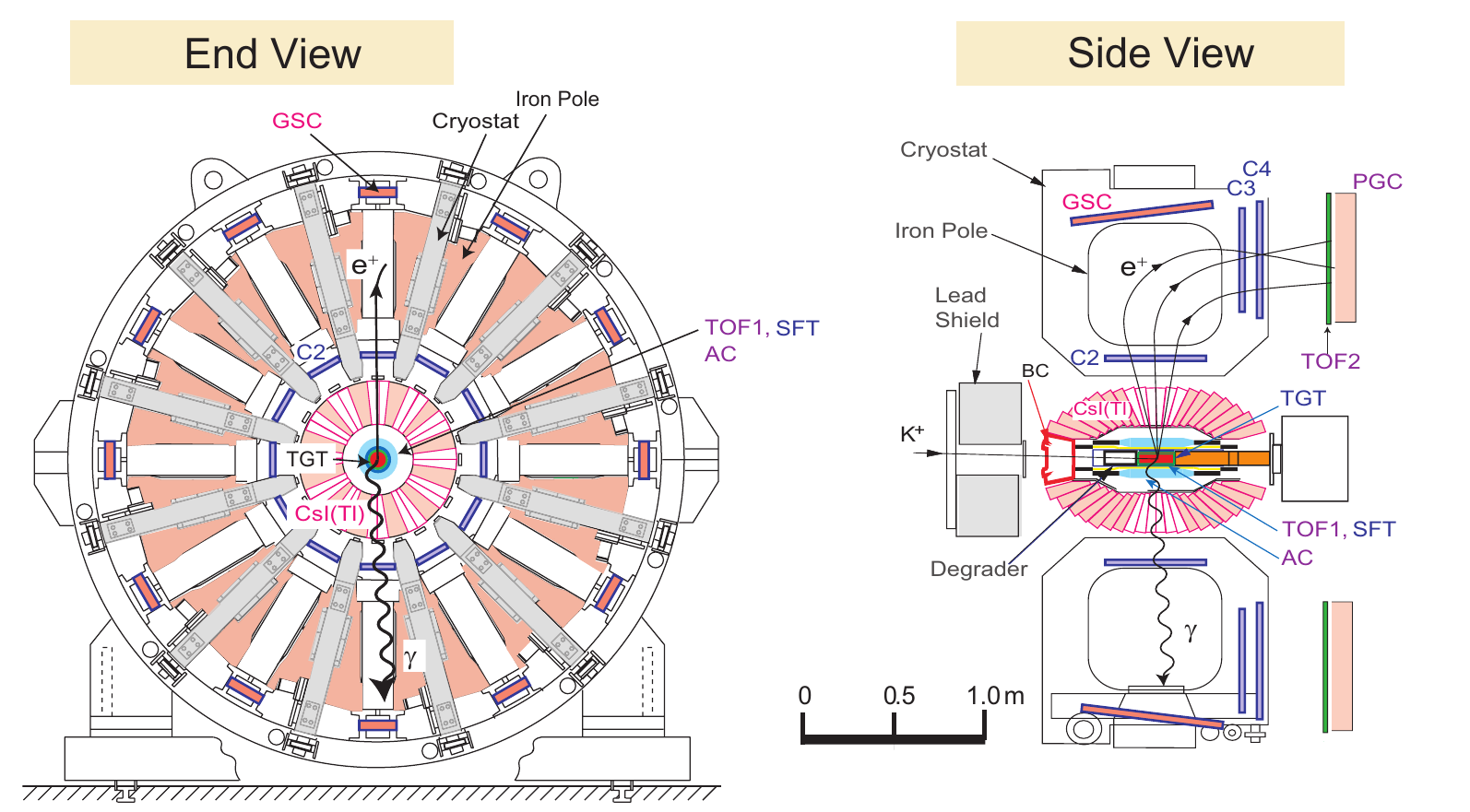}
  \end{center}
  \caption{TREK/E36 apparatus. General end and side views of the detector system.}
  \label{fig:e36apparatus}
 \end{figure}

 TREK/E36 was conducted at the K1.1BR kaon beamline in the hadron hall facility at J-PARC using a stopped $K^+$ beam in conjunction with a 12-sector superconducting toroidal spectrometer \cite{Abe:2003wi,Abe:2006de}. Figure \ref{fig:e36apparatus} shows the end and side view of the E36 detector apparatus. Details of the experimental method and detector configuration have been described in Ref.\cite{J-PARCE36:2021yvz}.
 
 The incoming $K^+$ was tagged with a Fitch \v{C}erenkov detector \cite{Abe:2003wi} in order to distinguish it from a $\pi^+$, before being slowed down by a BeO degrader and stopped in the active target (TGT). Charged $K^+$ decay products were tracked by using the spiral fiber tracker (SFT) \cite{Mineev:2016oew} which surrounded the target, and three multi-wire proportional chamber (C2, C3, C4) in each spectrometer sector. The C2, and C3/C4 were placed at the entrance and exit of the magnet gaps, respectively. To suppress $\mu^+/e^+$ mis-identification, redundancy was introduced into the particle identification detectors which was provided by time of flight detectors (TOF1,  TOF2), aerogel \v{C}erenkov, and lead glass \v{C}erenkov counters (PGC). A highly segmented and large acceptance CsI(Tl) calorimeter barrel which consisted of 768 crystals and covered about 70\% of $4\pi$ \cite{Dementev:2000hu} was used to identify and correct for structure dependent (SD) background events and also to search for light neutral bosons $A^{\prime}$.  The $A^{\prime}$ search necessitated that a $\mu^+$ is tracked in the gap, whose primary vertex lies  within the fiducial volume of the target (\textit{good gap event}) and two-clusters in the CsI calorimeter to detect the $e^+e^-$ pair, along with at least 3 TOF1 counters that registered a charged particle hit.

  \section{Analysis and Preliminary Results}
\begin{figure}[H]
  \begin{center}
   \includegraphics[scale=.31]{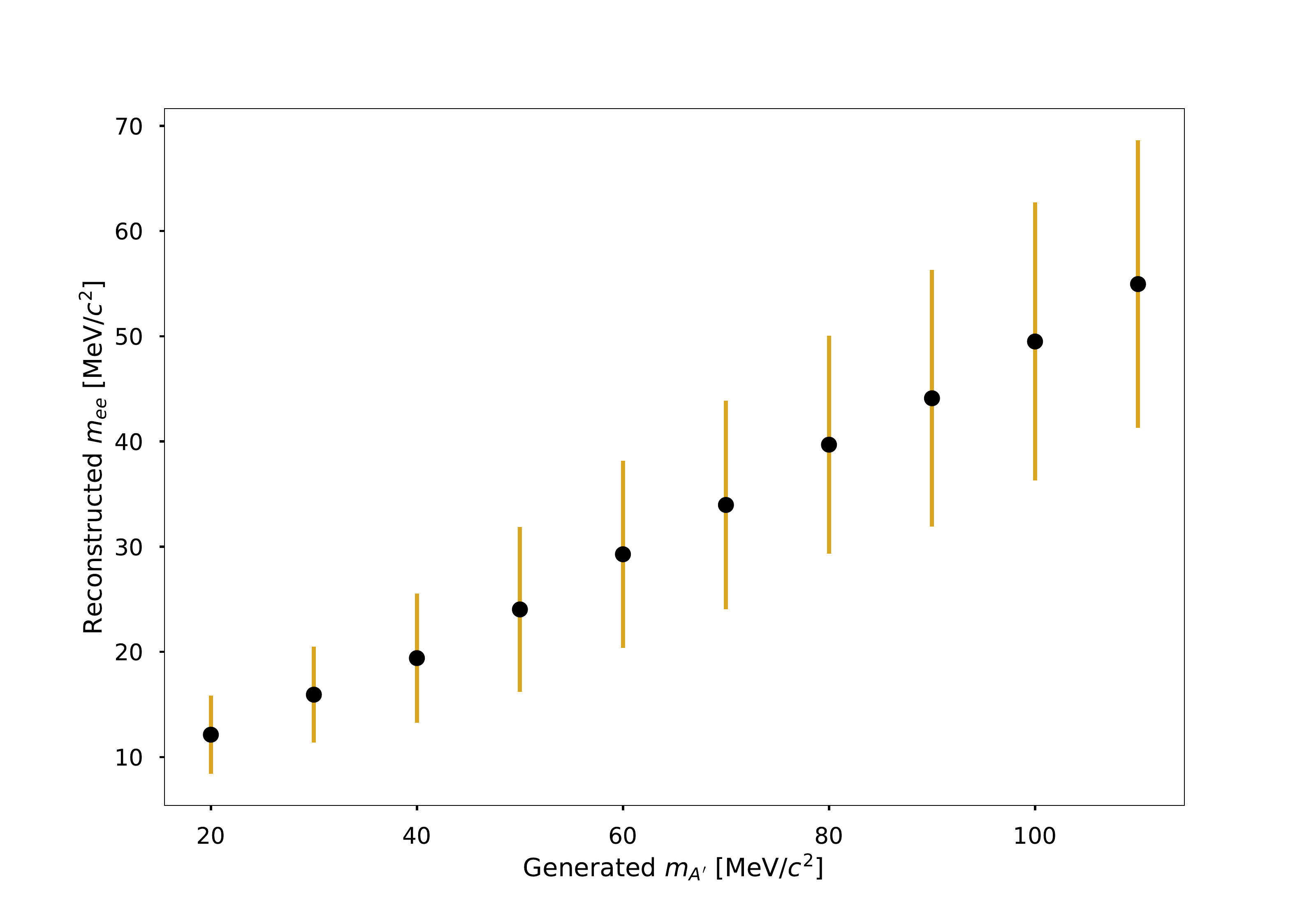}
  \end{center}
  \caption{MC reconstructed invariant mass $m_{ee}$ vs. generated $A^\prime$ mass. The horizontal bars represent a 2$\sigma$ search window for a given $A^\prime$ mass. \cite{Dongwi:2020ths}.}
  \label{fig:sigwind}
 \end{figure}

 The aforementioned $A^{\prime}$ signal search required a TOF1 multiplicity of of at least 3, implying that three charged particles have passed through the TOF1 counters. Specifically the $\mu^+$ is tracked in the magnet gap and the $e^+e^-$ form clusters in the CsI(Tl) calorimeter, and an invariant mass $m_{ee}$ distribution was reconstructed from the clusters. A Geant4 \textit{Monte Carlo} was used to generate and reconstruct $A^\prime$ mass $M_{A^\prime}$, whose reconstructed widths, corresponding to 2$\sigma$, were used as the search window in the $m_{ee}$ spectrum, shown in Figure \ref{fig:sigwind}\cite{Dongwi:2020ths}. 
 
 This analysis considered masses in the range of 20-110 MeV/$c^2$, in incremental step-sizes of 5 MeV/$c^2$. Several cut conditions in addition to the TOF1 multiplicity were employed in order to suppress the reducible background:
 
 \begin{enumerate}[label=\roman*., itemsep=0pt, topsep=0pt]
  \item a momentum endpoint cut as a function of the generated $M_{A^\prime}$
  \item a correlated angle cut between fired TOF1 counter and corresponding CsI(Tl) cluster 
  \item a muon mass squared $M_{\mu}^2$ cut for the charged particle in the gap.
 \end{enumerate}

\begin{figure}[H]
  \begin{center}
   \includegraphics[scale=0.52]{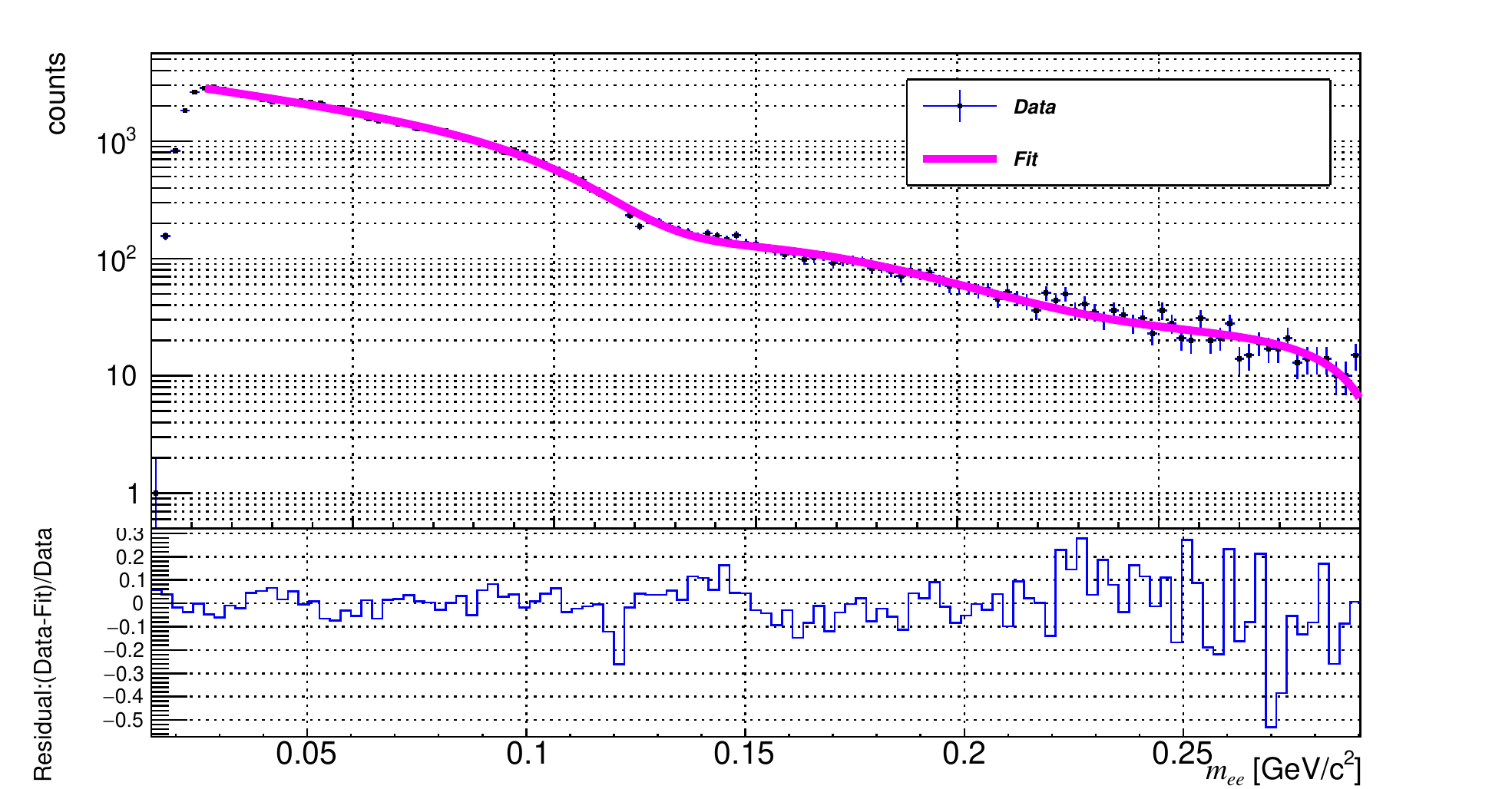}
  \end{center}
  \caption{Invariant mass $m_{ee}$ distribution with fit to data. The normalized residual is shown in the bottom plot.}
  \label{fig:fitInvmass}
 \end{figure}

 \noindent A polynomial fit to the resulting invariant mass $m_{ee}$ data can be seen in Figure 4 below. The bottom panel of Figure \ref{fig:fitInvmass} shows the residual, which is defined as the relative difference between data and fit, normalized to the data.
 
 The fit function from Figure \ref{fig:fitInvmass} and the total number of stopped kaons $N_K$ were used to extract the upper limits. The total number of stopped $K^+$ in the fiducial region of the active target was computed as follows
 
 \begin{align}
  N_K&=\frac{N_{\mu2}}{\mathcal{B}r(K_{\mu2})A_{\mu2}}\nonumber \\
  &=2.81\times10^9
  \label{eqn:numKaons}
 \end{align}

 \noindent where $N_{\mu2}$ is the number of tracked data candidates satisfying the $K_{\mu2}$ decay momentum, $A_{\mu2}$ is the acceptance fraction of the $K_{\mu2}$ muons evaluated with the MC simulations and $\mathcal{B}r(K_{\mu2})$ is the nominal branching ratio of the $K_{\mu2}$ decay mode. Upper limits at 95\% CL on the branching ratio of $Br(K^+\rightarrow\mu^+\nu_{\mu}A^\prime)$ for each $A^\prime$ mass corresponding to a $2\sigma$ limit for not observing the $A^\prime$ was computed for each aforementioned mass and corresponding search window of the invariant spectrum using the following relation
 
 \begin{equation}
  Br(K^+\rightarrow\mu^+\nu_{\mu}A^\prime)<
  \frac{2\sqrt{N_{\mu\nu ee}}}{N_{K}A_{A^\prime}\cdot LT}.
  \label{eqn:upperlim}
 \end{equation}

 \noindent where $N_{\mu\nu ee}$ is the number of candidate events within a given search window after applying all cuts to suppress the reducible background, $A_{A^\prime}$ is the fractional acceptance of the signal process $K^+\rightarrow\mu^+\nu_{\mu}A^\prime$, $A^\prime\rightarrow ee$, and LT is the livetime fraction for candidate events. The upper limit as a function of the invariant mass $m_{ee}$ is shown in Figure 6 (a). With $\sim20\%$ of the data analyzed, the upper limit was found to be at $\mathcal{O}(10^{-6})$. 

 \begin{figure}[H]
  \begin{center}
   \begin{tabular}{@{}c@{}}
    \includegraphics[scale=.47]{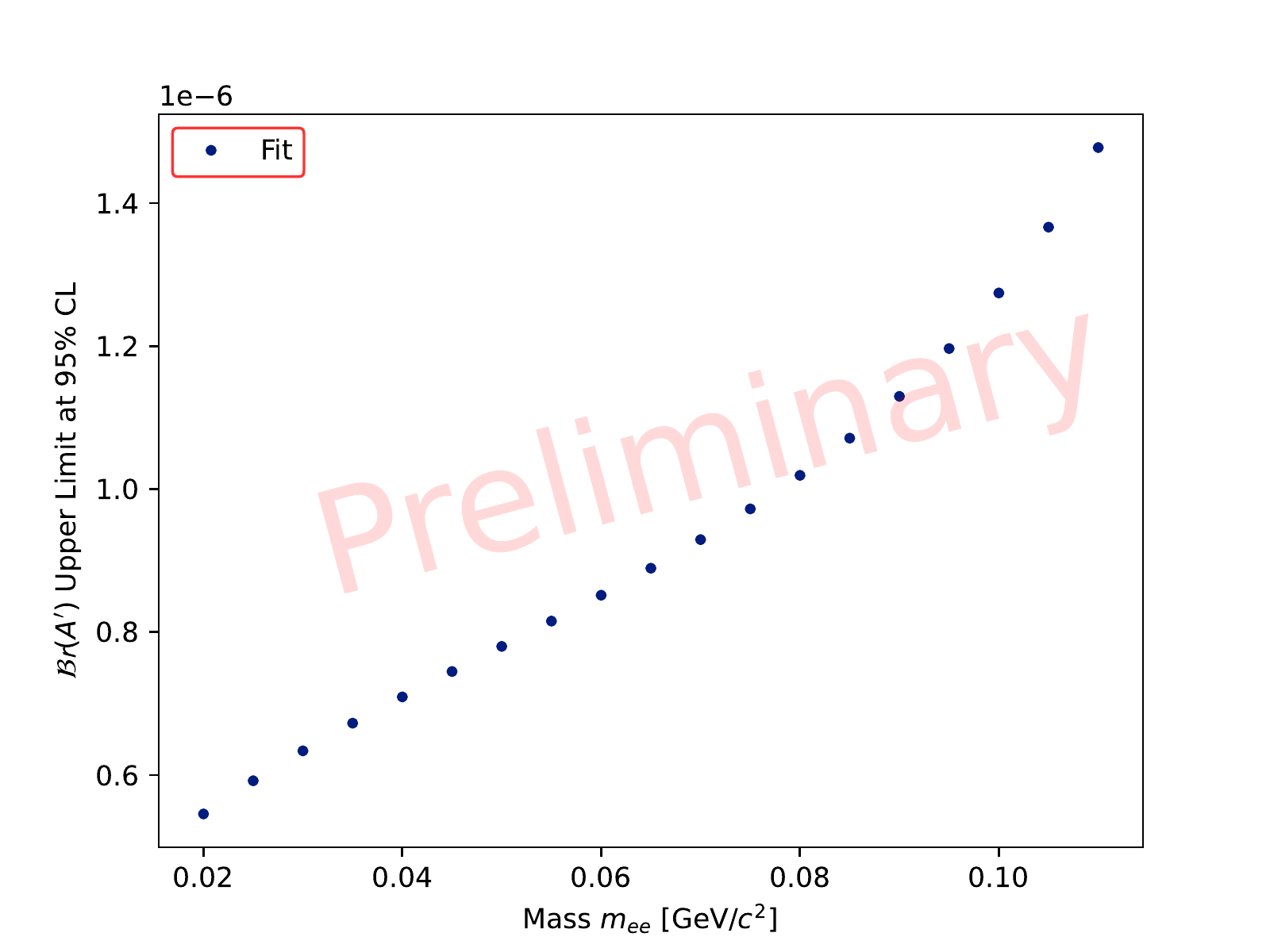} \\ (a)
   \end{tabular}
   \begin{tabular}{@{}c@{}}
    \includegraphics[scale=.47]{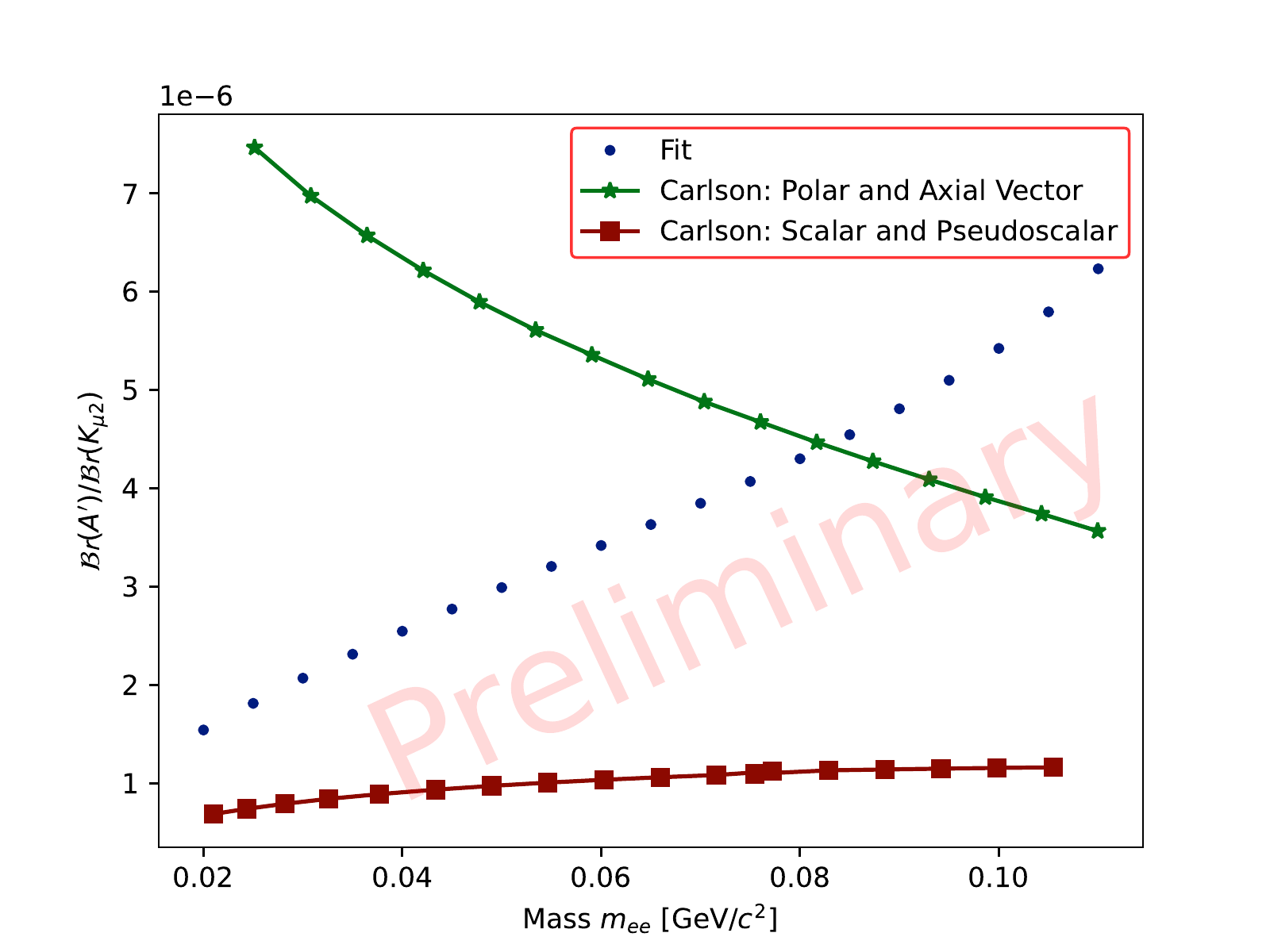} \\ (b)
   \end{tabular}
  \end{center}
   \caption{Upper limit extraction (a) as a function of reconstructed invariant mass $m_{ee}$. Data (blue) and theory comparisons (b) on particles with scalar and pseudoscalar (squares), and polar and axial vector (stars) couplings. Particles with both polar and axial vector couplings appear to be excluded for masses below 80 MeV/$c^2$.}
  \label{fig:uplim}
 \end{figure}

 Experimental limits for the Carlson model for new light neutral bosons with polar (scalar) and axial vector (pseudoscalar) couplings are $\mathcal{O}(10^{-6})$. Whereas Figure 6 (b), shows the comparison between data and theory prediction of particles with scalar and pseudoscalar (green), and polar and axial vector (orange) couplings \cite{Carlson:2012pc}. An $A^\prime$ particle with both polar and axial vector couplings, appears to be already excluded for masses below 80 MeV/$c^2$ with only $\sim20\%$ of the data analyzed. However, an $A^\prime$ with scalar and pseudoscalar couplings is not yet excluded. With further development in reducible background suppression, and more data analysis, we are hoping to enhance the sensitivity and experimental reach. Presently PID cuts to suppress $e/\mu/\pi$ mis-identification have not been applied. These cuts would further reduce the reducible background, and would enhance our sensitivity to particles with scalar and pseudoscalar couplings.

 \section*{References}
 \raggedright
 \bibliographystyle{iopart-num}
\bibliography{bibliography}

 \section*{Acknowledgements}
 This work has been supported in part by NSERC and NRC in Canada, the KAKEN-HI Grants-in-Aid for Scientific Research in Japan, the DOE and NSF in the US, and in Russia by the Ministry of Science and Technology.\\
 Prepared by LLNL under Contract DE-AC52--07NA23744.
 
 \newpage
Examples taken from published papers:
\medskip

\numrefs{99}
\item Kurata M 1982 {\it Numerical Analysis for Semiconductor Devices} (Lexington, MA: Heath)
\item Selberherr S 1984 {\it Analysis and Simulation of Semiconductor Devices} (Berlin: Springer)
\item Sze S M 1969 {\it Physics of Semiconductor Devices} (New York: Wiley-Interscience)
\item Dorman L I 1975 {\it Variations of Galactic Cosmic Rays} (Moscow: Moscow State University Press) p 103
\item Caplar R and Kulisic P 1973 {\it Proc. Int. Conf. on Nuclear Physics (Munich)} vol 1 (Amsterdam: 	North-Holland/American Elsevier) p 517
\item Cheng G X 2001 {\it Raman and Brillouin Scattering-Principles and Applications} (Beijing: Scientific) 
\item Szytula A and Leciejewicz J 1989 {\it Handbook on the Physics and Chemistry of Rare Earths} vol 12, ed K A Gschneidner Jr and L Erwin (Amsterdam: Elsevier) p 133
\item Kuhn T 1998 {\it Density matrix theory of coherent ultrafast dynamics Theory of Transport Properties of Semiconductor Nanostructures} (Electronic Materials vol 4) ed E Sch\"oll (London: Chapman and Hall) chapter 6 pp 173--214
\endnumrefs
\end{document}